\documentclass[11pt]{article}

\def\be{\begin{equation}}
\def\ee{\end{equation}}
\def\ba{\begin{eqnarray}}
\def\ea{\end{eqnarray}}
\def\dfrac{\displaystyle\frac}
\def\nn{\nonumber}
\def\lb{\label}
\def\bb{\bibitem}
\def\nn{\nonumber}
\def\A{{\cal A}}
\def\E{{\cal E}}
\def\C{{\cal C}}
\def\R{R_{(3)}}
\def\ol{\overline}
\def\dz{\partial_{\zeta}}
\begin{document}
\begin{titlepage}
\date{19 November 2015}
\title{\begin{flushright}\begin{small}    LAPTH-063/15
\end{small} \end{flushright} \vspace{1.5cm}
Axisymmetric multiwormholes revisited}
\author{G\'erard Cl\'ement \thanks{Email:
gclement@lapth.cnrs.fr} \\ \\
{\small LAPTh, Universit\'e Savoie Mont Blanc, CNRS,} \\
{\small 9 chemin de Bellevue,  BP 110, F-74941 Annecy-le-Vieux
cedex, France}}

\maketitle

\begin{abstract}
The construction of stationary axisymmetric multiwormhole solutions
to gravitating field theories admitting toroidal reductions to
three-dimensional gravitating sigma models is reviewed. We show
that, as in the multi-black hole case, strut singularities always
appear in this construction, except for very special configurations
with an odd number of centers. We also review the analytical
continuation of the multicenter solution across the $n$ cuts
associated with the wormhole mouths. The resulting Riemann manifold
has $2^n$ sheets interconnected by $2^{n-1}n$ wormholes. We find
that the maximally extended multicenter solution can never be
asymptotically locally flat in all the Riemann sheets.
\end{abstract}
\end{titlepage}

\section{Introduction}
Black holes attract each other, so stationary multi-black hole
configurations should not be possible in principle. Notwithstanding
this, stationary axisymmetric multi-black-hole solutions to
Einstein's equations were built long ago \cite{bachweyl} by linearly
superposing aligned Schwarzschild solutions. Such solutions are
possible only because of the presence of strut-like conical
singularities along the axis, which account for the extra balancing
forces \cite{israelkhan,letelier}. However balance without struts is
possible if both positive and negative charges are allowed, as
discussed in \cite{israelkhan,bns} where regular multi-center
solutions with an odd number of carefully chosen and positioned
sources were constructed.

Besides black hole solutions and solutions with naked singularities,
some gravitating field theories also admit spherically symmetric
Lorentzian wormhole solutions, the first known being due to Ellis
\cite{ellis} and Bronnikov \cite{kb}. As in the black hole case, it
should be possible to linearly superpose two or more aligned
wormhole solutions to obtain a stationary axisymmetric
multi-wormhole configuration. Indeed, this technique was applied
some time ago to the generation of axisymmetric multi-wormhole
solutions to five-dimensional general relativity \cite{gc84c} from
the Chodos-Detweiler (CD) spherically symmetric wormhole solution
\cite{CD82}. After reduction to four dimensions, the CD wormhole is
massless and electrically charged, so two such wormholes should
repel each other. Nevertheless, it was argued in \cite{gc84c} that
the strut singularities should be altogether absent in the
multi-wormhole case. A similar construction -- and a similar
argument -- was applied in \cite{gc84d} to the Bronnikov wormhole
solution to four-dimensional Einstein-Maxwell theory with a massless
phantom scalar field (coupled repulsively to gravitation), and
generalized in \cite{gc84e} to wormhole solutions with NUT charge to
the same theory.

Recently, the supercritically charged Reissner-Nordstr\"om-NUT
(RN-NUT) solution to the Einstein-Maxwell equations was shown to
correspond to a geodesically complete, traversable Lorentzian
wormhole \cite{wnut}. Due to the presence of a NUT charge, the
RN-NUT metric is not asymptotically Minkowskian. A possible cure for
this would be to consider it as a building block from which to
construct a NUT-anti-NUT, two-wormhole solution. The two monopole
NUT charges would compensate, to give a net dipole charge (angular
momentum) at infinity. A solution built up from two spherically
symmetric wormholes should be axisymmetric, so it should be possible
to construct it using the technique of \cite{bachweyl,gc84c}, if all
the charges associated with the second wormhole were opposite to
those of the first wormhole. The sum of the attractive forces
between opposite electric and magnetic charges and of the repulsive
forces between opposite masses and NUT charges would then lead,
because the RN-NUT wormholes are supercritically charged, to a net
attractive force, so that struts should be necessary for balance.

It therefore seems necessary to reconsider the construction of
stationary axisymmetric multiwormhole solutions and the issue of
strut singularities. In the next section we review the construction
of wormhole solutions to gravitating field theories as geodesic
solutions to their three-dimensional toroidal reductions. We then
revisit in Sect. 3 the construction of aligned multiwormhole
solutions from linear superpositions of harmonic functions centered
on the symmetry axis. We show why the naive regularity argument of
\cite{gc84c} fails, so that struts holding the wormholes in place
necessarily appear in this construction, except for certain
fine-tuned configurations with an odd number of centers. Analytical
continuation to a multi-sheeted Riemann manifold is discussed in
Sect. 4. We find that struts can never be avoided in the maximally
extended spacetime, which is always asymptotically conical in some
of the Riemann sheets. Our results are summarized in the concluding
section.

\setcounter{equation}{0}
\section{Wormholes from target space geodesics}
Consider a four-dimensional gravitating field theory, possibly
obtained by toroidal dimensional reduction from some
higher-dimensional theory, involving a set of gravity-coupled scalar
fields $\phi^a$ and $U(1)$ vector fields $A^I$. The stationary
sector of the theory may be further reduced to three dimensions by
assuming the generic stationary metric parametrization
 \be\lb{m}
ds^2 = -f(dt + \A_i dx^i)^2 + f^{-1}h_{ij}\,dx^i dx^j\,,
 \ee
where the various fields depend only on the three reduced space
coordinates $x^i$ ($i=1,2,3$). Solving the mixed $G_0^i = T_0^i$
Einstein equations enables to trade the Kaluza-Klein vector field
$\A_i$ for a three-dimensional scalar twist potential $\chi$, while
the four-dimensional $U(1)$ vector fields may similarly be traded
for pairs of scalar electric $v^I$ and magnetic $u^I$ potentials.
This reduction leads to the gravitating three--dimensional
$\sigma$-model action \cite{BMG}
\begin{equation}
S_{(3)} = \int d^3x\sqrt{|h|}\bigg[-\R +
G_{AB}(X)\partial_iX^A\partial_jX^Bh^{ij}\bigg]\,,
\end{equation}
where $\R$ is the Ricci scalar built out of the $h_{ij}$. The
equations of motion for the scalars $X^A=(f, \chi, v^I, u^I,
\phi^a)$
\begin{equation} \label{hm}
\frac{1}{\sqrt{h}}\partial_i\left(\sqrt{h}h^{ij}G_{AB}
\partial_j X^B\right)=0
\end{equation}
define a harmonic map from the three--space $\{x^i\}$ to the target
space or potential space $\{X^A\}$ with the line element
\begin{equation}\label{tar}
dS^2 = G_{AB}dX^AdX^B \,.
\end{equation}
In many cases of interest, this target space is a symmetric space or
coset (for instance $SU(2,1)/S[U(1,1)\times U(1)]$ in the
Einstein-Maxwell case, or $SL(3,R)/SO(2,1)$ in the case of
five-dimensional general relativity), which leads to fruitful
solution-generating techniques. The symmetric space property will
not be used in the present paper.

If now we assume that the coordinates $X^A$ depend on a single
potential function $\sigma(\vec{x})$, we have the freedom to choose
this potential to be harmonic \cite{neukra},
 \be\lb{harm}
\partial_i(\sqrt{h}h^{ij}\partial_j\sigma) = 0\,.
 \ee
The sigma-model field equations (\ref{hm}) then reduce to the
geodesic equations for the target space (\ref{tar}), which are first
integrated by
 \be
G_{AB}\frac{dX^A}{d\sigma}\frac{dX^B}{d\sigma} = 2\epsilon
 \ee
($\epsilon$ constant), while the three-dimensional Einstein
equations reduce to
 \be\lb{riccisig}
R_{(3)ij} = 2\epsilon\,\partial_i\sigma\partial_j\sigma\,.
 \ee
The target space metric is indefinite, the two gravitational
potentials $f$ and $\chi$ and the original scalar fields $\phi^a$
countributing positively to the metric (\ref{tar}) and the electric
and magnetic potentials $v^I$ and $u^I$ contributing negatively (the
signs are reversed in the case of phantom fields). Accordingly, the
geodesics fall in three classes depending on the value of
$\epsilon$: a timelike class ($\epsilon>0$), which includes black
hole solutions; a null class ($\epsilon=0$), which corresponds to
extremal black holes; and a spacelike class ($\epsilon<0$), which
includes wormhole solutions. Choosing in this last case the
convenient normalization $\epsilon=-1$, and assuming spherical
symmetry, the coupled equations (\ref{harm}) and (\ref{riccisig})
are solved by the harmonic function and reduced space metric
 \ba
&& \sigma = \arctan\left(\dfrac{b}{r}\right) \,,  \lb{sigell}\\
&& dl^2 \equiv h_{ij}dx^idx^j = dr^2 + (r^2+b^2)(d\theta^2 +
\sin^2\theta d\varphi^2) \lb{redell}\,.
 \ea

Note that the reduced three-space has the wormhole topology, the
coordinate $r$ varying in the whole real axis, with two asymptotic
regions $r=\pm\infty$, and a wormhole neck of radius $b$ at $r=0$.
This necessary condition for the four- or higher-dimensional metric
to correspond to a Lorentzian wormhole is not sufficient, in
addition the gravitational potentials $f$ and $\chi$ (as well as the
Kaluza-Klein potentials and moduli in the higher-dimensional case)
must also be regular for all $r$. The case of the CD wormhole
solution to five-dimensional general relativity is treated in
\cite{gc84c}. In the case of the RN-NUT wormhole solution to the
four-dimensional Einstein-Maxwell equations, the solution of the
target space geodesic equations yields for the complex Ernst
potentials $\E = f + i\chi -\ol{\psi}\psi$ and $\psi = v + iu$:
 \be
\E = \dfrac{1-\mu\tan\sigma}{1+\mu\tan\sigma}\,, \quad \psi =
\dfrac{\gamma\tan\sigma}{1+\mu\tan\sigma} \,, \lb{Epsisig}
 \ee
where the complex parameters $\mu$ and $\gamma$ of the geodesic
under consideration are related by
 \be
|\gamma^2| - |\mu^2| = 1\,.
 \ee
For the harmonic function (\ref{sigell}) this gives
 \be
f(r) = \frac{r^2+b^2}{(r+m)^2+n^2}\,, \quad \chi(r) =
\frac{-2nr}{(r+m)^2+n^2}\,,
 \ee
with
 \be
b\mu = m+in\,, \quad b\gamma = q - ip \quad (b^2 =
q^2+p^2-m^2-n^2)\,.
 \ee
The four-dimensional metric (\ref{m}) with NUT parameter $n$ then
corresponds to a wormhole with a neck of areal radius $n$ at $r=-m$.

\setcounter{equation}{0}
\section{Multicenter configurations}
More generally, assuming only axisymmetry of the three-dimensional
metric, which may be written in the Weyl form as
 \be\lb{weyl}
dl^2 = e^{2k}(d\rho^2 + dz^2) + \rho^2\,d\varphi^2\,,
 \ee
the harmonicity condition (\ref{harm}) on $\sigma(\rho,z)$ and
equation (\ref{riccisig}) read
 \ba
&& \rho^{-1}(\rho\sigma_{,\rho}){,\rho} + \sigma_{,z,z} = 0 \,, \lb{harma}\\
&& \dz k = -2\rho\,(\dz\sigma)^2\,, \lb{ksig}
 \ea
where we have combined the two cylindrical coordinates $\rho$ and
$z$ into the complex variable
 \be
\zeta = \rho + iz\,.
 \ee
Because of the linearity of (\ref{harma}), any number of spherically
symmetric solutions $\sigma_p$ of this equation may be superposed to
yield an axisymmetric solution $(\E,\,\psi,\,h_{ij})$ depending on
$\sigma = \sum_p\sigma_p$. In the case of the RN-NUT wormhole, the
resulting four-dimensional metric will be (\ref{m}), with the
reduced metric given by (\ref{weyl}) and (\ref{ksig}), and
 \be\lb{fAsig}
f = \frac{1+\tan^2\sigma}{|1+\mu\tan\sigma|^2}\,, \qquad \dz
\A_{\varphi} = 2i\,{\rm Im}\mu\,\rho\,\dz\sigma\,,
 \ee
so that the Kaluza-Klein potential $\A_{\varphi}$ will also be a
linear superposition.

The first step is to cast the spherically symmetric reduced metric
(\ref{redell}) in the Weyl form (\ref{weyl}). Following \cite{gc84c}
we first introduce a new radial coordinate $R>0$ related to $r$ by
 \be
r = R - \dfrac{b^2}{4R}\,,
 \ee
leading to the isotropic form of the solution
(\ref{sigell})-(\ref{redell})
 \ba
&& \sigma = 2\arctan\left(\dfrac{b}{2R}\right) \,, \\
&& dl^2 = \left(1+\dfrac{b^2}{4R^2}\right)^2\left[dR^2 + R^2\,
(d\theta^2 + \sin^2\theta d\varphi^2)\right]\lb{rediso}\,.
 \ea
The metric (\ref{rediso}) may then be transformed to the Weyl form
by the conformal Joukovski transformation
 \be\lb{zetachi}
\zeta = \chi + \frac{b^2}{4\chi}\,,
 \ee
where we have put
 \be
\chi = iRe^{-i\theta}\,.
 \ee
The harmonic function $\sigma$ and the Weyl metric function $k$ are
then given by the complex analytical continuation $m \to ib$ of the
formulas valid for the Schwarzschild solution \cite{weyl}
 \be\lb{sigketa}
\sigma = \arctan\left(\dfrac{b}{{\rm Re}\,\eta}\right)\,,\quad
e^{2k} = \frac{({\rm Re}\,\eta)^2 + b^2}{|\eta|^2}\,,
 \ee
with
 \be\lb{etazeta}
\eta^2 = (\zeta-b)(\ol\zeta+b)\,.
 \ee
The square root of (\ref{etazeta}) can be written explicitly as
 \be\lb{etachi}
\eta =
|\chi|^{-1}\left(\chi-\dfrac{b}2\right)\left(\ol\chi+\dfrac{b}2\right)\,,
 \ee
where $\chi(\zeta)$ is given by the inverse function of
(\ref{zetachi}),
 \be\lb{chizeta}
\chi = \frac12[\zeta + (\zeta^2 - b^2)^{1/2}]\,.
 \ee
This inverse function is bivalued, with two determinations $\chi^+$
and  $\chi^-$ corresponding to the two determinations of the square
root function, and defines a two-sheeted Riemann surface, the two
sheets (corresponding to the two sides of the wormhole) being
connected along the cut $C$ $(z=0,\,-b \le\rho\le b)$ (the wormhole
neck).

The superposition of two harmonic functions
$\sigma_1(\rho,z-a_1;b_1)$ and $\sigma_2(\rho,z-a_2;b_2)$ (we assume
in the following $a_1 < a_2$) yields the harmonic function
 \be
\sigma = \sigma_1 + \sigma_2 = \arctan\left(\dfrac{b_2{\rm
Re}\,\eta_1 + b_1{\rm Re}\,\eta_2}{{\rm Re}\,\eta_1\,{\rm
Re}\,\eta_2 - b_1b_2}\right)\,,
 \ee
with
 \be\lb{etapchi}
\eta_p^2 = (\zeta_p-b_p)(\ol\zeta_p+b_p) \qquad (\zeta_p = \rho +
i(z-a_p))\,,
 \ee
and the Weyl metric function
 \be
k = k_1 + k_2 + 2k_{12}\,,
 \ee
where $k_p$ are the Weyl functions (\ref{sigketa}) for the
respective harmonic functions $\sigma_p$, and $k_{12}$ is given by
the analytical continuation of the formula given in
\cite{bachweyl,israelkhan,letelier,bns}
 \be\lb{k12}
e^{2k_{12}} = \left\vert\dfrac{2\eta_1\ol\eta_2 +
(\zeta_1-b_1)(\ol\zeta_2-b_2) +
(\zeta_2+b_2)(\ol\zeta_1+b_1)}{2\eta_1\eta_2 +
(\zeta_1-b_1)(\ol\zeta_2+b_2) +
(\zeta_2-b_2)(\ol\zeta_1+b_1)}\right\vert,.
 \ee
In the special case of the superposition of a wormhole centered at
($\rho=0$, $z=-a$) and an antiwormhole centered at ($\rho=0$,
$z=a$), $b_1=-b_2=b$ and $a_1=-a_2=-a$.

The integration constant in (\ref{k12}) has been chosen so that the
boundary condition
 \be\lb{bound}
k(\infty) = 0\,,
 \ee
which ensures that the reduced three-space is asymptotically
Euclidean, is satisfied. It was argued in \cite{gc84c} that, because
from (\ref{ksig}) $k(\rho,z)$ is constant on the symmetry axis
$\rho=0$, this also ensures that the regularity condition (absence
of conical singularities)
 \be\lb{reg}
k(0,z) = 0
 \ee
is satisfied for all $z$. However this argument overlooked the fact
that the functions $\chi_p(\zeta)$, and therefore also the functions
$\eta_p(\zeta)$ given by (\ref{etapchi}), are discontinous along the
cuts $C_p$. Indeed these discontinuities are similar to the disk
discontinuities observed in the superposition of a ring and a
homogeneous field \cite{hoen} or a Chazy-Curzon particle
\cite{letelier}. In the present multiwormhole case, the metric can
be smoothly continued through the disks into other Riemann sheets
(see next section), so that these discontinuities are not associated
with matter sources. However they do lead to step-function jumps in
$k_{12}(0,z)$ when a cut is crossed. It is clear from
(\ref{etapchi}) that, on the symmetry axis, the values of the two
determinations of the function $\eta_p(\rho,z)$ just above and just
below the cut $\C_p$ are related by $\eta_p^+(0,a_p) =
-\eta_p^-(0,a_p)$. While $k_{12}(0,z) = 0$ for $z<a_1$ or $z>a_2$
(below the two cuts or above the two cuts), a careful computation to
order $\rho^2$ leads to the value (the analytical continuation of
the result of \cite{israelkhan,letelier,bns})
 \be\lb{strut}
e^k(0,z) = e^{2k_{12}}(0,z) = \dfrac{(b_1+b_2)^2 +
(a_1-a_2)^2}{(b_1-b_2)^2 + (a_1-a_2)^2}\,,
 \ee
on the segment $a_1<z<a_2$ between the two cuts. This means that
this segment corresponds to a strut with positive (for two charges
$b_1$ and $b_2$ of the same sign) or negative (for two charges  of
opposite signs) tension keeping the two wormhole mouths apart. In
the case of the RN-NUT wormhole with $b_1+b_2=0$, this strut
coincides with the Dirac-Misner string connecting the two oppositely
charged wormhole mouths.

The above construction is easily generalized to the superposition of
$n$ harmonic functions,
 \be\lb{sigmasum}
\sigma = \sum_{p=1}^n\sigma_p\,,
 \ee
leading to the Weyl function
 \be\lb{ksum}
k = \sum_{p=1}^n k_p + 2\sum_{q=2}^n\sum_{p=1}^{q-1}k_{pq}\,.
 \ee
As in the multi-black-hole case, a strutless, regular multiwormhole
configuration can be achieved in the case of an odd number of
suitably charged and positioned wormholes. We consider only the
example of three equally spaced wormholes with $(a_1,a_2,a_3) =
(-a,0,a)$, and $(b_1,b_2,b_3) = (b,c,b)$. Between the cuts $\C_1$
and $\C_2$, $k_{23}(0,z) = 0$, so that
 \be
e^k(0,z) = e^{2k_{12}}(0,z)e^{2k_{13}}(0,z)\,.
 \ee
By reason of symmetry, the result is the same between $\C_2$ and
$\C_3$, so that the solution is regular if
 \be\lb{strutless}
\dfrac{(b+c)^2+a^2}{(b-c)^2+a^2}\;\dfrac{b^2+a^2}{a^2} = 1\,,
 \ee
which is solved by
 \be
a^2 = - \dfrac{b(b+c)^2}{b+4c}\,,
 \ee
provided $b(b+4c)<0$. For instance, a configuration with net
monopole charges equal to zero, $2b+c=0$, will be in equilibrium for
an intercut distance $a = b/\sqrt7$.

\setcounter{equation}{0}
\section{Analytical continuation}

Now we discuss the analytical continuation of the multiwormhole
solution across the $n$ cuts $C_p$. To this end we denote the two
possible determinations of the inverse Joukovski function by
 \be\lb{chizetapm}
\chi_p^{\pm} = \frac12[\zeta_p \pm (\zeta_p^2 - b_p^2)^{1/2}]\,.
 \ee
Analytical continuation across the cut $C_p$ corresponds to the
replacement
 \be
\chi_p^{\pm} \longrightarrow \chi_p^{\mp} =
\frac{b_p^2}{4\chi_p^{\pm}}\,.
 \ee
Accordingly, the complex function $\eta_p^{\pm}$ and the real
harmonic function $\sigma_p^{\pm}$ (defined modulo $\pi$) are
replaced by
 \be
\eta_p^{\mp} = -\eta_p^{\pm}\,, \quad  \sigma_p^{\mp} = -
\sigma_p^{\pm}\,.
 \ee
The maximally extended three-dimensional manifold has $2^n$ Riemann
sheets interconnected by $2^{n-1}n$ wormholes \cite{gc84c}.

Consider first the case $n=2$, and label the four sheets $(++)$,
$(+-)$, $(-+)$, and $(--)$. The respective functions $\sigma(\zeta)$
and $k_{12}(\zeta)$ are
 \ba
\sigma^{++} &=& - \sigma^{--} = \sigma_1 + \sigma_2\,, \quad
k_{12}^{++} = k_{12}^{--} = k_{12}\,, \nn\\
\sigma^{+-} &=& - \sigma^{-+} = \sigma_1 - \sigma_2\,, \quad
k_{12}^{+-} = k_{12}^{-+}\,,
 \ea
with
 \be\lb{k12pm}
e^{2k_{12}^{+-}} = \left\vert\dfrac{-2\eta_1\ol\eta_2 +
(\zeta_1-b_1)(\ol\zeta_2-b_2) +
(\zeta_2+b_2)(\ol\zeta_1+b_1)}{-2\eta_1\eta_2 +
(\zeta_1-b_1)(\ol\zeta_2+b_2) +
(\zeta_2-b_2)(\ol\zeta_1+b_1)}\right\vert
 \ee
(the functions $k_p$, which are even in $\eta_p$, are unchanged). In
the sheet $(+-)$ or $(-+)$, the effective charges are now
$(b_1,-b_2)$ or $(-b_1,b_2)$, so that inter-wormhole forces which
are repulsive (attractive) in the first sheet $(++)$ become
attractive (repulsive) in those sheets.

By continuation through one or the other cut, we expect that the
Weyl function $k^{+-}(\rho,z)$ will now satisfy the regularity
condition (\ref{reg}) on the segment of the symmetry axis
$a_1<z<a_2$ between the two cuts, and take the value (\ref{strut})
on the complementary segments $z<a_1$ and $z>a_2$, which now act as
struts counterbalancing the net attractive (repulsive) force between
the effective charges. It follows that $k(0,\pm\infty)$ no longer
vanishes, so that the boundary condition (\ref{bound}) cannot be
satisfied. Indeed, direct computation of (\ref{k12pm}) in the limit
$(\rho \to \infty,\, z \to \infty)$ yields the value
 \be\lb{k+-}
k^{+-}(\infty) = 2k_{12}^{+-}(\infty) = \ln\dfrac{(b_1+b_2)^2 +
(a_1-a_2)^2}{(b_1-b_2)^2 + (a_1-a_2)^2}\,,
 \ee
in accordance with the value (\ref{strut}) on the struts $z<a_1$ and
$z>a_2$. Thus the reduced three-space cannot be asymptotically
Euclidean in all the Riemann sheets. If, as we have assumed, the
Euclidean boundary condition (\ref{bound}) holds in the sheets
$(++)$ and $(--)$, then the reduced three-metric is asymptotically
conical in the sheets $(+-)$ and $(-+)$.

In spite of this discouraging result, one could hope that the
strutless three-center configurations discussed in the preceding
section would survive analytical continuation, or at the very least
that some three-center configuration could be found to be
asymptotically locally flat in all the Riemann sheets. However the
answer turns out to be negative. Consider a generic three-center
configuration obeying the Euclidean boundary condition (\ref{bound})
in the sheets $(+++)$ and $(---)$. Then in the other sheets we find
 \ba
&& k^{-++}(\infty) = k^{+--}(\infty) = 2[k^{+-}_{12}(\infty) +
k^{+-}_{13}(\infty)]\,, \nn \\
&& k^{+-+}(\infty) = k^{-+-}(\infty) = 2[k^{+-}_{23}(\infty) +
k^{+-}_{12}(\infty)]\,,  \\
&& k^{++-}(\infty) = k^{--+}(\infty) = 2[k^{+-}_{13}(\infty) +
k^{+-}_{23}(\infty)]\,. \nn
 \ea
These can simultaneously vanish only if $k^{+-}_{12}(\infty) =
k^{+-}_{13}(\infty) = k^{+-}_{23}(\infty)=0$, implying from
(\ref{k+-}) than two of the three charges $b_i$ vanish, so that the
solution reduces to the one-wormhole solution. This argument, which
generalizes to the case of $n$ centers, shows that any multi-center
reduced three-metric constructed fom the superposition (\ref{ksum})
is necessarily asymptotically conical in some sheets of its maximal
analytic extension.

\section{Conclusion}
We have generalized the construction of stationary axisymmetric
multiwormhole solutions given in \cite{gc84c} to gravitating field
theories admitting toroidal reductions to three-dimensional
gravitating sigma models. We have shown that -- as in the
multi-black hole case -- strut singularities, necessary to restore
the otherwise unequal balance between attractive and repulsive
forces, always appear in this construction, except for very special
configurations with an odd number of centers.

We have also reviewed the analytical continuation of the multicenter
solution across the cuts associated with the wormhole mouths.
Contrary to naive intuition, the resulting Riemann manifold is not
two-sheeted, but has $2^n$ sheets interconnected by $2^{n-1}n$
wormholes, where $n$ is the number of cuts. Some interwormhole
forces change sign when going from one sheet to another adjacent
sheet, so that the strut configurations are different. We have shown
that, as a consequence, the maximally extended multicenter solution
can never be asymptotically locally flat in all the Riemann sheets.

\section*{Acknowledgments}
I thank Dmitry Gal'tsov for discussions and a careful reading of the
manuscript. I also acknowledge discussions with Adel Bouchareb.

\end{document}